# Growth and Characterization of GaAs Nanowires on Carbon Nanotube Composite Films: Towards Flexible Nano-Devices

Parsian K. Mohseni[a], Gregor Lawson[b], Christophe Couteau[c], Gregor Weihs[c], Alex Adronov*[b] and Ray R. LaPierre*[a]

[a] Center for Emerging Device Technologies, Department of Engineering Physics, McMaster University, Hamilton, Ontario, L8S 4L7, Canada
[b] Department of Chemistry and the Brockhouse Institute for Materials Research, McMaster University, Hamilton, Ontario, L8S 4L7, Canada
[c] Institute for Quantum Computing, University of Waterloo, Waterloo, Ontario, N2L 3G1, Canada

* Authors to whom correspondence should be addressed
E-mail: lapierr@mcmaster.ca, adronov@mcmaster.ca

## Abstract

Poly(ethylene imine) functionalized carbon nanotube thin films, prepared using the vacuum filtration method, were decorated with Au nanoparticles by in-situ reduction of $HAuCl_4$ under mild conditions. These Au nanoparticles were subsequently employed for the growth of GaAs nanowires (NWs) by the vapor-liquid-solid process in a gas source molecular beam epitaxy system. The process resulted in the dense growth of GaAs NWs monolithically integrated across the entire surface of the single-walled nanotube (SWNT) films. The NWs, which were orientated in a variety of angles with respect to the SWNT films, ranged in diameter between 20 to 200 nm, with heights up to 2.5 μm. Transmission electron microscopy analysis of the NW-SWNT interface indicated that NW growth was initiated upon the surface of the nanotube composite films. Photoluminescence characterization of a single NW specimen showed high optical quality. Rectifying, asymmetric current-voltage behavior was observed from contacted NW ensembles and attributed to the core-shell pn-junction within the NWs.

Throughout the last decade, carbon nanotubes and semiconductor nanowires (NWs) have reached the forefront of nanometer-scale materials science by offering novel device architectures and applications that take advantage of their unique material properties[1]. Specifically, III-V compound semiconductor NWs have been the focus of extensive research based on their role as the active element within devices such as single-electron transistors[2] and light-emitting diodes[3]. Single NW-based photovoltaic devices have also attracted a great deal of interest in recent academic endeavors[4, 5]. Use of NWs in solar cells promises improved photoconversion efficiency through the enhanced carrier collection processes offered by core-shell architectures[6], while reducing cost and improving functionality through integration with a wide range of substrates.

Research surrounding carbon nanotube (CNT) thin films has also created considerable interest due, in part, to their potential application in the fabrication of electronic devices. CNT films have been shown to exhibit resistivity and optical transparency approaching that of conventional transparent conductive materials such as Indium-Tin-Oxide (ITO),[7] while also displaying superior flexibility[8] and impressive environmental resistance. Moreover, CNT films may be fabricated with relative ease using a variety of low-cost room temperature techniques including spin coating[9], casting[8, 10-12], spraying[13, 14], Langmuir-Blodgett film formation[15], layer-by-layer (LbL) assembly with polyelectrolytes[16-18], electrodeposition[19], and vacuum filtration[20-24]. CNT thin films therefore offer a potential low-cost alternative to ITO in electronic devices such as chemical sensors[25-28], light-emitting diodes[29-32] and solar cells[7, 33-35], specifically in applications where a high degree of flexibility is desirable. In addition to their impressive conductivity, CNTs are known to exhibit exceptional thermal stability[36-38] and have been reported to behave as p-type Ohmic contacts within GaN-based LEDs, with contact resistances lower than conventional metallic contacts[39]. This unique combination of properties renders CNT thin films excellent candidates for use as flexible conducting substrates for growth of semiconductor NWs. Although CNTs have been demonstrated as templates for the assembly of various supplementary nanostructures[40-42], little consideration has been made to date with regard to their integration with single crystal semiconductor NWs, over macroscopic scales.

Our group recently reported a simple method for the preparation of single-walled nanotube (SWNT) thin films decorated with Au nanoparticle clusters[43]. In this work, SWNTs functionalized with highly branched poly(ethylene imine) (PEI) were shown to exhibit impressive aqueous solubility, allowing for the formation of homogeneous thin films by vacuum filtration. Incubation of PEI-functionalized SWNT films in $HAuCl_4$ solution resulted in the formation of Au nanoparticle clusters in high density upon the film surface. Here, we report the employment of this procedure in the preparation of conductive SWNT films functionalized with Au-nanoparticles that are demonstrated as suitable substrates for the assembly of III-V semiconductor NWs, according to the vapor-liquid-solid (VLS) mechanism[44-46], in a gas source molecular beam epitaxy (GS-MBE) system. Whereas the heteroepitaxial growth of NWs on Si and ITO-coated substrates[47] over large areas has been established, the use of CNT composites as the growth surface has hitherto remained unexplored. This work represents the first union between the characteristic optoelectronic benefits of GaAs NWs and the inherent flexibility and conductivity offered by CNT films.

SWNTs used in this work were purchased from Carbon Nanotechnologies, Inc. (Houston, TX). The SWNTs were produced by the HiPco process and were used as received. All other reagents and solvents were obtained from commercial suppliers and used without any prior treatment. SWNT films decorated with Au nanoparticles were prepared using the previously described procedure[43]. To obtain nanoparticles having diameters suitable for NW growth, films were submersed in 0.5 mM $HAuCl_4$ solution for one minute at room temperature. After films had been washed to remove traces of $HAuCl_4$, they were dried under nitrogen, removed from the Teflon membrane support and transferred, Au-particle side up, to Si wafers, having dimensions of approximately 1 $cm^2$. The films were subsequently subjected to a rapid thermal annealing (RTA) treatment at 550 °C for 10 minutes, under nitrogen ambient.

Prior to growth initiation, the samples were heated to the growth temperature of 550 °C under an $As_2$ overpressure. In the GS-MBE system, group III species (Ga) were supplied as monomers from a heated solid elemental effusion cell while the group V species were supplied as dimers ($As_2$) from a hydride ($AsH_3$) gas cracker operating at 950 °C. NW growth was initiated by opening the Ga effusion cell shutter, preset to yield a nominal two-dimensional film growth rate of 1 μm/hr under a constant V/III flux ratio of 1.5. For the purposes of this study, two distinct NW architectures were studied (groups A and B). Group A NWs were strictly composed of nominally undoped GaAs, while group B NWs were composed of pn-junction core-shell heterostructures. For the group A sample, nominally undoped NWs were grown for 30 minutes. In the case of the group B sample, a primary GaAs layer was grown over a 15 minute period, nominally n-doped with Te to a concentration of $10^{18}$ $cm^{-3}$. Next, a secondary GaAs layer was deposited for 15 minutes, nominally p-doped with Be to a concentration of $10^{18}$ cm$^{-3}$. Doping concentrations were calibrated using previous depositions of GaAs epilayers on GaAs (100) substrates. For both group A and group B NWs, the growth was terminated, after a total period of 30 minutes, by closing the shutter to the Ga cell and allowing the samples to cool down from the growth temperature under an As$_2$ overpressure.

The orientation, morphology, and density of the as-grown samples were investigated using a JEOL JSM-7000 scanning electron microscope (SEM) and a Carl Zeiss SMT NVision 40 dual-beam microscope, in secondary-electron mode. The latter, equipped with focused ion beam (FIB) capability, was used in the preparation of lamellae for cross-sectional transmission electron microscopy (TEM) using a Philips CM12 microscope and a JEOL 2010F high-resolution transmission electron microscope (HR-TEM). For analysis of a single NW specimen, the as-grown samples were subjected to a 60 second ultra-sonication process in a small volume of de-ionized water after which suspended NWs were deposited on a holey carbon-coated copper grid. Similarly, single NWs were dispersed on an oxidized-Si substrate for micro-photoluminescence (μ-PL) characterization in a continuous flow helium cryostat at 7 K. Excitation was provided using a laser centered at 532 nm at a power of 80 μW. Excitation and μ-PL collection were achieved through an objective allowing for an excitation spot of roughly 2 μm. PL was resolved by a 75 cm grating spectrometer and detected by a liquid nitrogen-cooled Si charge-coupled device camera.

Figure 1 shows planar SEM images of SWNT-Au composite films before and after the RTA annealing treatment. Covalent modification of SWNTs by PEI-functionalization results in the conversion of numerous carbon atoms within the framework of nanotubes to sp³ hybridization. Thermal annealing of PEI-functionalized SWNT films induces decomposition of the functional groups attached to the surface of the nanotubes and results in partial restoration of the nanotube electronic structure and conductivity[48]. Furthermore, an annealing treatment ensures removal of volatile organic matter that might otherwise contaminate the MBE chamber used for semiconductor NW growth. Following this treatment, irregularly shaped Au-nanoparticle clusters were found to re-form into smaller, discrete nanoparticles having an average diameter of approximately 38 nm as shown in Figure 1(b).

SEM analysis in Figure 2 indicated that the as-grown samples from group A exhibited dense NW growth over the entire span of the Au-functionalized SWNT films. Similar results were obtained from the group B sample. In comparison to the vertically oriented growth of GaAs NWs on GaAs (111)B or Si (111) substrates[49, 50], NWs grown on SWNT substrates were observed to be oriented in a variety of angles with respect to the growth surface. This can be understood by considering the well-established premise that NWs preferentially assemble in the thermodynamically favorable <111> or <0001> directions, for zincblende and wurtzite crystal structures, respectively[49, 51]. However, in the case of the present study, the growth surface lacks long range spatial periodicity, which is necessary for epitaxial orientation of the NWs, due to the random stacking arrangement of single-walled CNTs in the composite film.

Predominantly, both group A and group B NWs were of characteristic tapered or "pencil-shaped" morphologies, with heights up to 2.5 μm and average full-width at half-maximum diameters of roughly 100 nm. The tapered NW morphology is attributed to sidewall diffusion-limited radial growth, as previously reported[50]. The simultaneous occurrence of layer-by-layer radial deposition and Au-nanoparticle-based axial growth results in the core-shell architecture of the NWs[49]. A two-dimensional GaAs film, formed concurrently with the NWs, is evident in Figure 2(b) above the SWNT substrate. Prior growths on single crystalline Si substrates indicated two-dimensional film thicknesses of 120 nm[49]. In comparison, the present growth resulted in deposition of a thicker film layer of approximately 450 nm, probably as a consequence of shorter adatom diffusion lengths on the rough surface of the SWNT films.

Figure 2(c) shows a tilted SEM view of an area specifically manipulated to demonstrate the flexibility of a SWNT film containing NWs. It is particularly remarkable that the NWs studied in this project maintained their structural integrity after bending, contorting, rolling, and folding of the underlying flexible substrate, over macroscopic scales.

Structural analysis of a single NW from the group A sample was carried out by TEM analysis, as shown in Figure 3. Figure 3(a) shows a TEM image of a NW that is representative of the structure and morphology of practically all NWs grown on the CNT films. Figure 3(b) reveals a magnified image of the identical NW in Figure 3(a). Contrast stripes intersecting the NWs indicated the presence of intermittent stacking faults. A selective area diffraction pattern obtained on the <2-1-10> zone axis of a defect-free NW segment, shown in the inset of Figure 3(a), confirmed a wurtzite crystal structure with wires growing along the <0001> direction. Consistent with previous experiments involving homo- and hetero-epitaxially grown NWs[49, 52], the stacking faults studied in the present case appeared as atomic layers arranged in a zincblende structure, amongst defect-free wurtzite segments. The Au-nanoparticle at the NW tip provides evidence for growth according to the VLS mechanism.

A point of interest is the nature of the growth surface. Due to the simultaneous deposition of a GaAs film during NW assembly, the NW/CNT interface becomes buried during the growth process. To investigate the NW/CNT interface, thin lamellae were prepared via FIB, allowing for the analysis of small cross-sectional windows where the internal NW structure, GaAs film, and CNT-composite substrate can be examined in a single specimen. In Figure 4, a TEM image is shown of a lamella removed from the as-grown sample. The coating surrounding the NW in Figure 4, is simply a carbon layer deposited during the sample preparation, as a protective envelope. Stacking faults can be seen throughout the entire length of the NW including the portion buried within the 2-D GaAs film. This observation leads to the argument that the initial NW nucleation process occurred at the CNT/Au interface. Hence, the GaAs film grows simultaneously between the NWs, but appears to play a negligible role in the NW growth process. However, due to the intimate contact between the NWs, 2-D film, and CNTs, the role of the planar GaAs layer is of importance to the electrical conduction pathways of this system as discussed later.

Micro-photoluminescence studies were conducted on single NWs placed on $SiO_2$ substrates. A typical μ-PL spectrum is shown in Figure 5 for a group A NW. Here, a single peak with 12.5 meV linewidth is obtained at 7 K, centered at 1.51 eV. The inset of Figure 5 shows a plot of the experimental shift in μ-PL peak energy with increasing temperature. The expected temperature dependence of the bandgap is plotted in the inset as a solid line according to the Varshni curve of bulk GaAs[49]. The agreement between the measured values and the Varshni expression indicates that the PL emission may be attributed to band-related recombination transitions within the undoped GaAs NW[52]. Agreement between the measured values and the expected trend illustrates the high crystallinity, purity, and optical quality of the NWs, making these materials promising for use in optoelectronic device applications.

The group A and B samples were further processed for electrical characterization. The main intent in the processing procedure was to ensure intimate electrical contact with the Au-capped NW tips, while avoiding a possible short-circuit pathway through the planar growth region. SEM images of the sample, at various processing steps employed in the device fabrication, are shown in Figure 6. First, the entire sample was coated with a $SiO_x$ layer (step a) formed through plasma-enhanced chemical vapor deposition (PE-CVD). This layer provides a conformal insulating shell across all NWs and the surface of the GaAs film, approximately 100 nm in thickness. Next, a polymer layer (S-1808 photoresist) was spin-cast over the oxide-coated NWs (step b). After a 1 min oxygen-plasma reactive ion etching (RIE) treatment, the top-most layers of the spin-on-polymer were removed to expose the oxidized NW tips (step c). At this point, the sample underwent a wet etch in a 10:1 buffered HF solution to remove the thin $SiO_x$ cap at the NW tips (step d). A second, more prolonged, RIE treatment occurred next, etching the remaining photoresist from the surface of the NWs. As seen in Figure 6(e), after this processing step, the planar GaAs layer and the body of the NWs remained insulated by the oxide layer, while the NW tips are exposed. Thus, once a Ti/Pt/Au top contact layer was deposited (step f), the intended pathway for electron flow is through the NWs to the SWNT film.

To study the electrical behavior of the fabricated devices and to establish the role of Be and Te as dopants, a Ti/Pt/Au layer was deposited by electron-beam evaporation on both group A (undoped GaAs NWs) and group B (pn-junction core-shell NWs) samples. Prior to measurement, the samples were treated at 400 °C for 30 seconds, to produce an Ohmic contact at the contact/NW interface. Figure 7 shows the current-voltage (I-V) characteristics of a bare SWNT film (post RTA), a device fabricated using group A NWs, and a device fabricated using group B NWs, under an applied bias of -2 V to 2 V.

It is worthy to note, first, that the SWNT film exhibited a conductive behavior prior to growth. From the I-V curve (square data points), SWNT films after being subjected to an annealing process revealed a sheet resistance of roughly 68 $\Omega$/sq (resistivity of $6.81 \times 10^{-3}$ $\Omega \cdot$cm). This value is in agreement with SWNT composite film resistances previously reported[8, 20, 53] and is comparable to sheet resistances measured in commercial ITO films[8]. Rectification at high biases is attributed to Schottky barriers formed between the CNT film and probe tips [54]. The undoped NWs (group A samples, circular data points) demonstrated a high resistivity of roughly 5900 $\Omega \cdot$cm. In contrast, the curve obtained from sample B containing pn-junction NWs (triangular data points) exhibited asymmetric rectification. The observation of diode-type behavior in group B devices, while not in group A devices, demonstrated the rectifying properties exhibited by the pn-junction NWs. This behavior opens the door to possible applications of these hybrid device architectures in a variety of applications, including photovoltaics, light-emitting diodes, and sensors.

In summary, a novel material combination involving epitaxially grown GaAs NWs on Au-functionalized SWNT composite films is reported. Au-nanoparticles on the surface of the SWNT sheets act as atomic sinks, collecting gas-phase adatoms supplied in a GS-MBE system and accommodating NW growth according to the VLS mechanism. As-grown NWs were oriented in a variety of angles on the SWNT surface and grew along the [0001] direction with wurtzite crystal structure. TEM analysis conducted on the NW/SWNT interface confirmed initiation of growth on the SWNT surface. Micro-PL characterization of a single NW specimen confirmed high optical quality. Electrically contacted pn-junction NWs exhibited rectifying behaviour, while undoped NWs showed high resistivity. Thus, a proof-of-concept potential for an emerging class of optoelectronic devices monolithically integrated with a conductive, flexible, and low-cost substrate as an alternative to conventional ITO is demonstrated. Similarly, CNT films are expected to accommodate VLS growth of alternative compound semiconductor material groups. Future efforts will focus on large-scale core-shell NW-based photovoltaics on carbon-nanotube composite fabric.

**Acknowledgements**

Financial support for this work was provided by the Natural Science and Engineering Research Council (NSERC) of Canada, the Canadian Foundation for Innovation (CFI), the Ontario Innovation Trust (OIT), the Premier's Research Excellence Award, and the Ontario Centres of Excellence. The authors gratefully acknowledge Julia Huang and Fred Pearson for assistance with FIB and HR-TEM.

**Figure Captions**

**Figure 1: Planar** SEM image of SWNT-Au composite film (a) before and (b) after annealing treatment. Bright spots indicate Au particles.

**Figure 2:** (a) Planar (b) cross-sectional and (c) tilted view SEM images of as-grown samples containing GaAs NWs on CNT composite films.

**Figure 3:** (a) TEM image of single NW. Inset shows SAD pattern, indicative of wurtzite structure and <0001> growth direction. (b) Magnified TEM view of stacking faults visible along NW lengths.

**Figure 4:** Cross-sectional TEM image of interfaces between NW, GaAs film, and CNT substrate. The extension of stacking faults within the GaAs film, localized to the lateral extent of the visible NW, indicates NW growth from the CNT surface.

**Figure 5:** Single NW μ-PL spectrum at 7 K. Inset plots the measured PL peak energy with increasing temperature as compared to the theoretical bulk GaAs Varshni curve.

**Figure 6:** Processing steps in device fabrication: (a) PE-CVD deposition of 100 nm $SiO_x$ layer to coat NWs and planar growth layer. (b) Coating of oxidized NWs with spin-on polymer (S1808 photoresist) layer. (c) Partial $O_2$-plasma RIE of polymer layer for exposure of oxidized NW tips. (d) Buffered HF wet etch treatment for the exposure of bare NW tips.  (e) Full $O_2$-plasma RIE for the removal of remaining polymer layer.  (f) Deposition of Ti/Pt/Au top contact, followed by RTA treatment.

**Figure 7:**  I-V characteristics of purified CNT film, prior to NW deposition (squares), fabricated devices containing undoped NWs (circles), and fabricated devices containing pn-junction NWs (triangles).  Asymmetric rectification in group B devices is indicative of diode-type behaviour from the pn-junction NWs.

**Figure 1:**

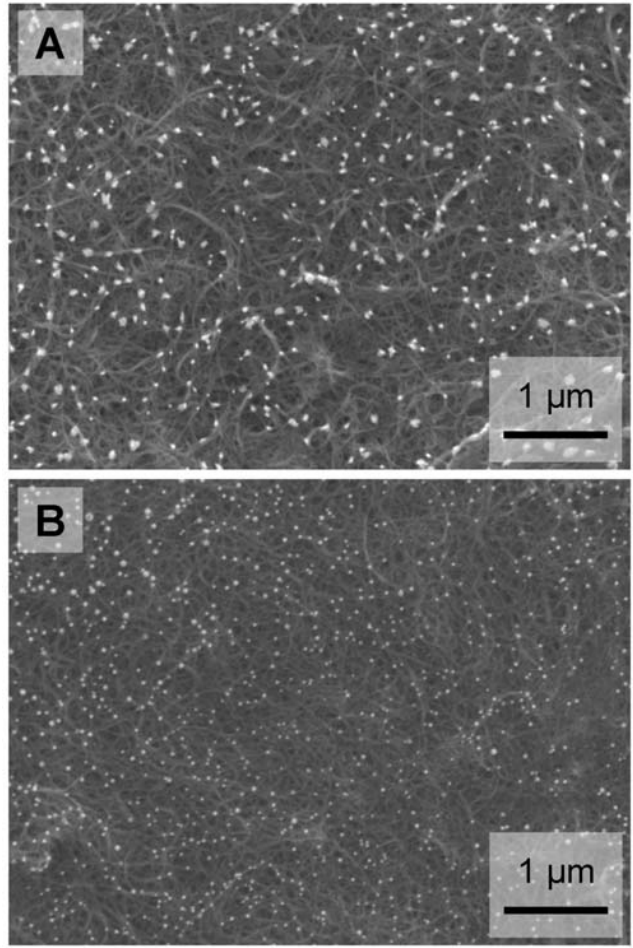

**Figure 2:**

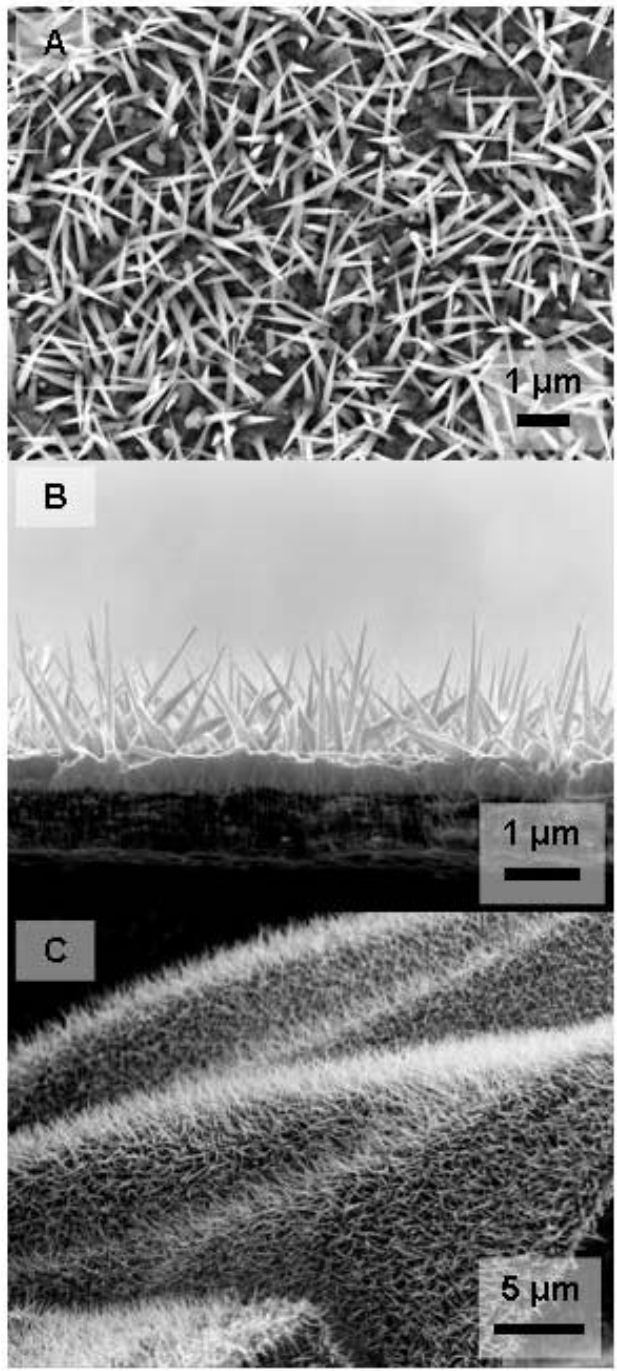

**Figure 3:**

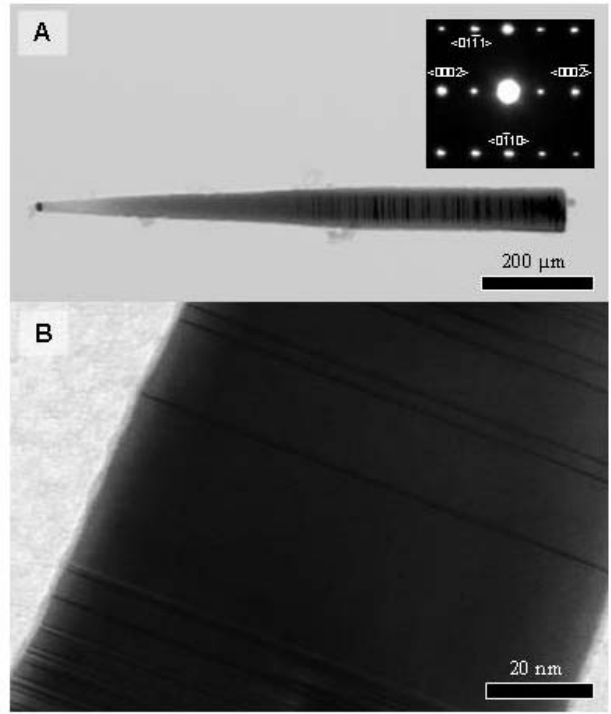

**Figure 4:**

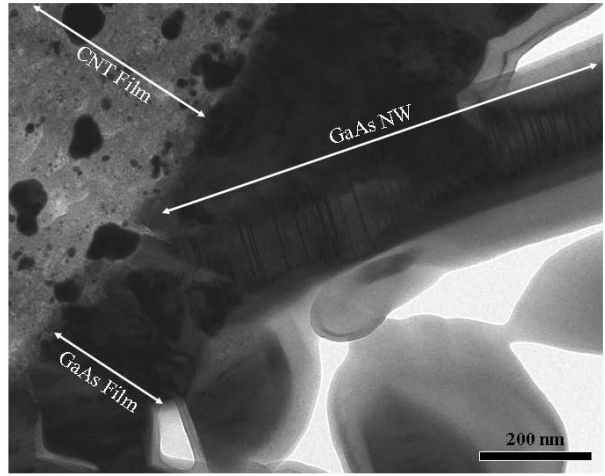

**Figure 5:**

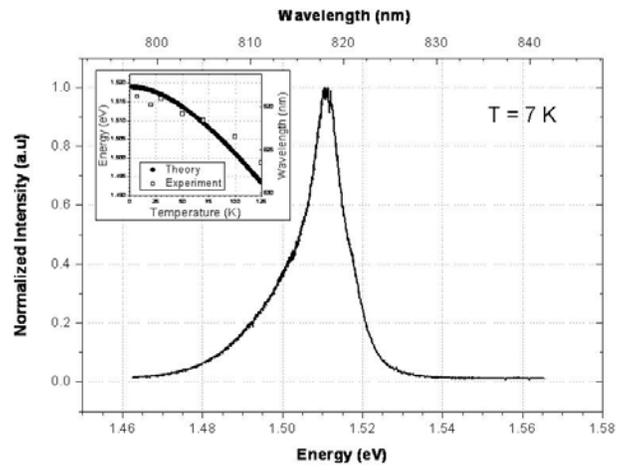

**Figure 6:**

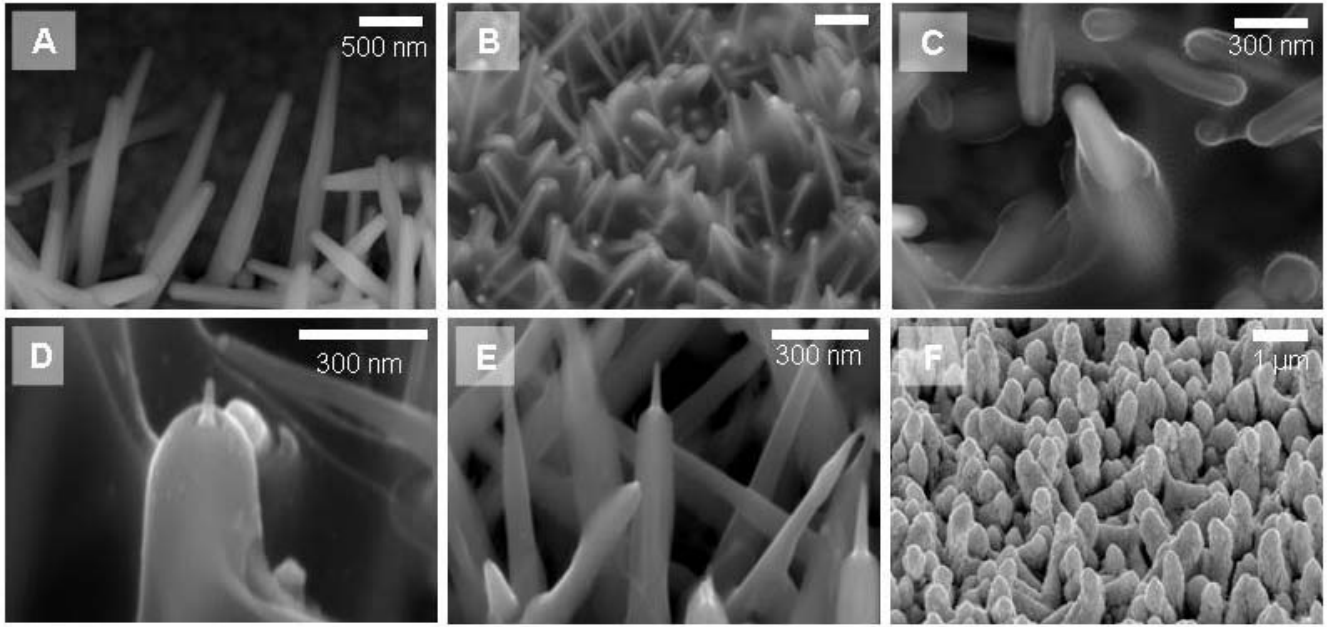

**Figure 7:**

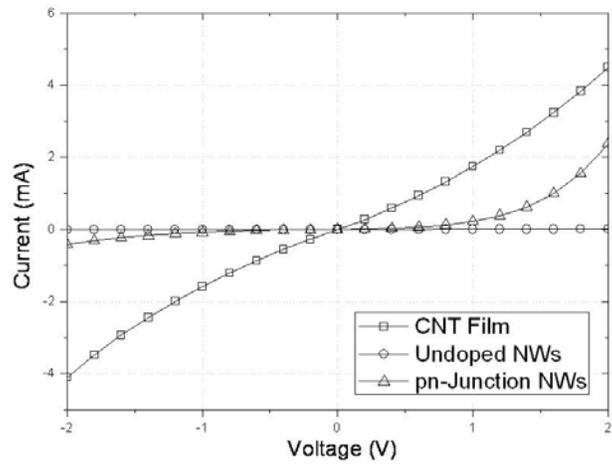